\documentclass[RNAAS]{aastex62}
\usepackage{comment}
\usepackage{color}
\usepackage{ulem}

\newcommand{\cplus}{[C\,{\sc ii}]}

\def\kms    {km\,s$^{-1}$}
\def\purple#1 {{\textcolor{purple}{#1}}\ }
\def\red#1 {{\textcolor{red}{#1}}\ }
\def\new#1 {{\bf #1 }}
\def\blue#1 {{\textcolor{blue}{#1}}\ }
\begin{document}

\title{ADF22: Blind detections of \cplus\ line emitters shown to be spurious}

\correspondingauthor{Natsuki~H.~Hayatsu}
\email{natsuki.hayatsu@utap.phys.s.u-tokyo.ac.jp}

\author[0000-0003-2112-1306]{Natsuki~H.~Hayatsu}
\affiliation{Department of Physics Graduate School of Science, The University of Tokyo, 7-3-1 Hongo, Bunkyo, Tokyo 113-0033, Japan}

\author[0000-0001-5118-1313]{R.\,J.~Ivison}
\author{Paola~Andreani}
\affiliation{European Southern Observatory, Karl-Schwarzschild-Str.~2, 85748 Garching, Germany}

\author{Hideki~Umehata}
\affiliation{RIKEN Cluster for Pioneering Research, 2-1 Hirosawa, Wako-shi, Saitama 351-0198, Japan}

\author{Yuichi~Matsuda}
\affiliation{National Astronomical Observatory of Japan, Osawa 2-21-1, Mitaka, Tokyo 181-8588, Japan}

\author{Naoki~Yoshida}
\affiliation{Department of Physics Graduate School of Science, The University of Tokyo, 7-3-1 Hongo, Bunkyo, Tokyo 113-0033, Japan}

\author{Kotaro~Kohno}
\author{Bunyo~Hatsukade}
\affiliation{Institute of Astronomy, Graduate School of Science, The University of Tokyo, 2-21-1 Osawa, Mitaka, Tokyo 181-0015, Japan}

\author{Akio~K.~Inoue}
\affiliation{Department of Physics, School of Advanced Science and Engineering, Waseda University, 3-4-1, Okubo, Shinjuku, Tokyo 169-8555, Japan}

\author{Yoichi~Tamura}
\author{Tutomu~T.~Takeuchi}
\affiliation{Division of Particle and Astrophysical Science, Graduate School of Science, Nagoya University, Furo-cho, Chikusa-ku, Nagoya, Aichi 464-8602, Japan}

\author{Seiji~Fujimoto}
\affiliation{Institute for Cosmic Ray Research, The University of Tokyo, Kashiwa, Chiba 277-8582, Japan}

\author[0000-0002-2419-3068]{Minju~M.~Lee}
\affiliation{Max-Planck-Institut für extraterrestrische Physik (MPE), Giessenbachstr., D-85748 Garching, Germany)}

\author{Tohru~Nagao}
\affiliation{Research Center for Space and Cosmic Evolution, Ehime University, 2-5 Bunkyo-cho, Matsuyama, Ehime 790-8577, Japan}

\author{Yiping~Ao}
\affiliation{Purple Mountain Observatory and Key Laboratory for Radio Astronomy, Chinese Academy of Sciences, 8 Yuanhua Road, Nanjing 210034, People's Republic of China}


\section{} 
We report Atacama Large Millimetre/submillimeter Array (ALMA) Cycle-5 follow-up observations of two candidate \cplus\ emitters at $z\approx 6$ in the ALMA deep field in SSA\,22 (ADF22). The candidates were detected blindly in a Cycle-2 ALMA survey covering $\approx 5$\,arcmin$^2$, with a single tuning, along with two CO lines associated with galaxies at lower redshifts.  Various tests suggested at least one of the two $\ge 6$-$\sigma$ \cplus\ candidates should be robust \citep{hayatsu2017}. Nevertheless, our new, deeper observations recover neither candidate, demonstrating a higher contamination rate than expected. The cause of the spurious detections is under investigation but at present it remains unclear.

\section{Observation and Analysis}

The data -- from Cycles 2 and 5 -- were analysed using Common Astronomy Software Application ({\sc casa}) version 5.1.1 \citep{mcmullin2007}. There were no significant differences between the new reduction and the original \citet{hayatsu2017} calibration using version 4.3.1.

\noindent
\textbf{Cycle-2 Observations:}
ADF22 was observed in ALMA band 6 in 2014 June and 2015 April (2013.1.00162.S, PI: H.~Umehata) using 33--36 12-m antennas in the C34-2 and C34-4 configurations with a precipitable water vapour (PWV) of 0.3--1.3\,mm.  The observations consisted of nine execution blocks and 103 pointings, with a typical on-source time per pointing of 4.5\,min. The correlator used time division mode (TDM), yielding four 1.875-GHz spectral windows (SPWs) with central frequencies of  254, 256, 270 and 272\,GHz, as detailed in \citet{umehata2017}.  

The blind line search described in \citet{hayatsu2017} resulted in the detection of two emission lines, at 253.79 and 269.92\,GHz, with J2000 positions:  (RA, Dec) = ($22^{\rm h} 17^{\rm m} 37.43^{\rm s}$, $+00^\circ 17' 10''.7$) and ($22^{\rm h} 17^{\rm m} 31.95^{\rm s}$, $+00^\circ 18' 20''.3$), respectively. The lines were seen in two independent subsets of the data \citep{hayatsu2017}.  The median r.m.s.\ values in the `dirty' (uncleaned) cube were around 0.8\,mJy\,beam$^{-1}$ at the original spectral resolution, $\approx 36$\,km\,s$^{-1}$. A known continuum source, ADF22.4, was detected with consistent positions and flux densities in the same pointing as ADF22-LineA \citep{umehata2017}, in both Cycles 2 and 5. 

\noindent
\textbf{Cycle-5 Observations:}
Observations to confirm the blind detections of the two \cplus\ emission-line candidates were undertaken in 2017 April (2017.1.00602.S, PI: N.\,H.~Hayatsu) in array configuration, C43-3, where the PWV was 1.4\,mm. On-source integration times were $\approx 11$\,min, and using frequency division mode (FDM), with $4 \times 1920$ dual-polarisation channels over a bandwidth of 7.5\,GHz, with four 1.875-GHz SPWs.  Single fields centred on the targets were observed at frequencies of 251.284 and 261.183\,GHz for ADF22-LineA and ADF22-LineB, respectively. 

The median noise value is $\approx 0.77$ \,mJy\,beam$^{-1}$ at $\approx  36$\,km\,s$^{-1}$ spectral resolution for both datasets, similar to the Cycle-2 data; the angular resolution was around $1''.0 \times 0''.6$.

\section{Results and Discussion}

\subsection{Comparison of data from Cycles 2 and 5}

Fig.~1 shows the Cycle-2 and 5 spectra of the line candidates, with a velocity resolution of 36\,\kms.  We also plot atmospheric transmission and r.m.s.\ noise.  Within the $\approx 200$-km\,s$^{-1}$ velocity range around the line centres, the emission-line features seen in Cycle-2 are not reproduced in our Cycle-5 observations.

From our Cycle-5 data we conclude that the emission-line candidates reported by \citet{hayatsu2017} were spurious; we cannot fully rule out the possibility of a transient line emitter, although the consistent detections in subsets of the data from 2014 and 2015 make a transient line emitter unlikely. Likely the smoothing used in our earlier work enhanced non-Gaussian noise and caused an under-estimate of the contamination rate.  Further discussion regarding the false detection rate and completeness in ALMA data, using mock observational data, will be described in Hayatsu\,et\,al.\,(in\,prep). The Cycle-2 data are undergoing a stage-3 quality-assurance process (QA3) to search for technical issues with the data.

\subsection{Implications for the \cplus\ luminosity function}

Our failure to recover either of the \cplus\ candidates allows us to set upper limits on the \cplus\ luminosities, $L_{\rm [CII]}$, of $3\sigma <2 \times 10^8$\,L$_\odot$.  

The resulting limit on the \cplus\ luminosity function (LF), together with the lower limit from \citet{swinbank2012} and the estimate for the local Universe from \citet{hemmati2017}, means the LF can evolve by 2--$3\times$ from $z = 0$--6; it cannot evolve by more than 10$\times$. 

\begin{figure*}
	\begin{center}
	\includegraphics[trim=0 0 0 0, width=160mm]{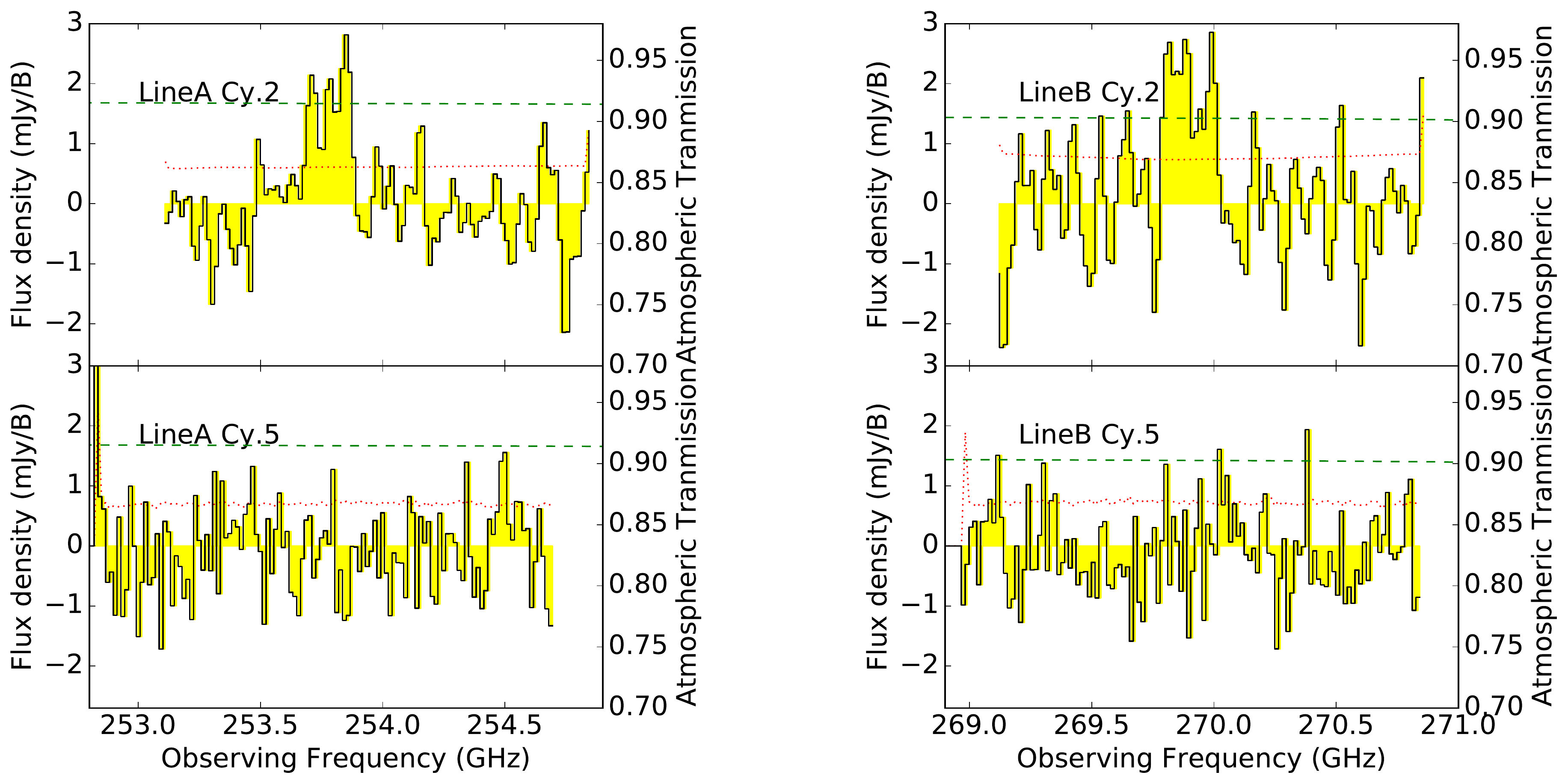}
  	\caption{Spectra of the ADF22-LineA (left) and ADF22-LineB (right) from the Cycle-2 (top) and Cycle-5 (bottom) observations, at a velocity resolution of $\approx 36$\,\kms.  The  green dashed line shows the atmospheric transmission (right axis) and the red dotted line shows the r.m.s.\ noise.  The emission lines seen in the Cycle-2 data are not recovered by the Cycle-5 observations.
  	}
	\label{figure2}
	\end{center}
\end{figure*}

\section{Acknowledgements}

This note makes use of the following ALMA data: ADS/JAO.ALMA\#2013.1.00162.S and ADS/JAO.ALMA\#2017.1.\\00602.S.  ALMA is a partnership of ESO (representing its member states), NSF (USA) and NINS (Japan), together with NRC (Canada) and NSC and ASIAA (Taiwan) and KASI (Republic of Korea), in cooperation with the Republic of Chile.  The Joint ALMA Observatory is operated by ESO, AUI/NRAO and NAOJ.


\begin{thebibliography}{}
\bibitem[Carilli, \& Walter(2013)]{carilli2013} Carilli, C.~L., \& Walter, F.\ 2013, \araa, 51, 105
\bibitem[Hayatsu et al.(2017)]{hayatsu2017} Hayatsu, N.~H., Matsuda, Y., Umehata, H., et al.\ 2017, \pasj, 69, 45 
\bibitem[Hemmati et al.(2017)]{hemmati2017} 
Hemmati, S., Yan, L., Diaz-Santos, T., et al.\ 2017, \apj, 834, 36 
\bibitem[McMullin et al.(2007)]{mcmullin2007} 
McMullin J.~P., Waters B., Schiebel D., Young W., Golap K., 2007, ASPC, 376, 127 
\bibitem[Swinbank et al.(2012)]{swinbank2012} 
Swinbank A.~M., et al., 2012, MNRAS, 427, 1066 
\bibitem[Umehata et al.(2017)]{umehata2017} 
Umehata, H., Tamura, Y., Kohno, K., et al.\ 2017, \apj, 835, 98
\bibitem[Williams, de Geus, \& Blitz(1994)]{williams1994} 
Williams J.~P., de Geus E.~J., Blitz L., 1994, ApJ, 428, 693 
\end{thebibliography}
\end{document}